\documentclass{article}

\usepackage{arxiv}

\usepackage[utf8]{inputenc} % allow utf-8 input
\usepackage[T1]{fontenc}    % use 8-bit T1 fonts
\usepackage{hyperref}       % hyperlinks
\usepackage{url}            % simple URL typesetting
\usepackage{booktabs}       % professional-quality tables
\usepackage{amsfonts}       % blackboard math symbols
\usepackage{nicefrac}       % compact symbols for 1/2, etc.
\usepackage{microtype}      % microtypography
\usepackage{lipsum}		% Can be removed after putting your text content
\usepackage{graphicx}
\usepackage{natbib}
\usepackage{doi}
\usepackage{dcolumn}% Align table columns on decimal point
\usepackage{bm}% bold math
\usepackage{siunitx}
\usepackage{authblk}

\title{On-Demand Single-Electron Source via Single-Cycle Acoustic Pulses}

\author[1,2]{Shunsuke Ota}
\author[3]{Junliang Wang}
\author[3]{Hermann Edlbauer}
\author[2]{Yuma Okazaki}
\author[2]{Shuji Nakamura}
\author[2]{Takehiko Oe}
\author[4]{\\Arne Ludwig}
\author[4]{Andreas D. Wieck}
\author[3]{Hermann Sellier}
\author[3]{Christopher B\"auerle}
\author[2]{Nobu-Hisa Kaneko}
\author[1]{Tetsuo Kodera}
\author[2,5]{Shintaro Takada}

\affil[1]{Department of Electrical and Electronic Engineering, Tokyo Institute of Technology, Tokyo 152-8550, Japan}
\affil[2]{National Institute of Advanced Industrial Science and Technology (AIST), National Metrology Institute of Japan (NMIJ), 1-1-1 Umezono, Tsukuba, Ibaraki 305-8563, Japan}
\affil[3]{Univ. Grenoble Alpes, CNRS, Grenoble INP, Institut N\'eel, 38000 Grenoble, France}
\affil[4]{Lehrstuhl f\"{u}r Angewandte Festk\"{o}rperphysik, Ruhr-Universit\"{a}t Bochum, Universit\"{a}tsstra\ss e 150, 44780 Bochum, Germany}
\affil[5]{Persent address: Department of Physics, Graduate School of Science, Osaka University, Toyonaka, Osaka 560-0043, Japan}
\affil[ ]{\textit {Corresponding author: takada@phys.sci.osaka-u.ac.jp}}

\renewcommand{\shorttitle}{On-Demand Single-Electron Source via Single-Cycle Acoustic Pulses}

\hypersetup{
pdftitle={On-Demand Single-Electron Source via Single-Cycle Acoustic Pulses},
pdfauthor={Shunsuke Ota},
}

\begin{document}
\maketitle

\begin{abstract}
	Surface acoustic waves (SAWs) are a reliable solution to transport single electrons with precision in piezoelectric semiconductor devices. Recently, highly efficient single-electron transport with a strongly compressed single-cycle acoustic pulse has been demonstrated. This approach, however, requires surface gates constituting the quantum dots, their wiring, and multiple gate movements to load and unload the electrons, which is very time-consuming. Here, on the contrary, we employ such a single-cycle acoustic pulse in a much simpler way – without any quantum dot at the entrance or exit of a transport channel - to perform single-electron transport between distant electron reservoirs. We observe the transport of a solitary electron in a single-cycle acoustic pulse via the appearance of the quantized acousto-electric current. The simplicity of our approach allows for on-demand electron emission with arbitrary delays on a \SI{}{ns} time scale. We anticipate that enhanced synthesis of the SAWs will facilitate electron-quantum-optics experiments with multiple electron flying qubits.
\end{abstract}

% keywords can be removed
% \keywords{First keyword \and Second keyword \and More}

\section{INTRODUCTION}
Surface acoustic waves (SAWs) are mechanical waves that propagate along a material surface and accompany electric field modulation in piezoelectric materials. This property allows for exquisite spatial and temporal control of the local environment of electrons in solid systems. This control makes SAWs an interesting and promising method for electron transport in electron-quantum optics \cite{Hermelin2011,McNeil2011,Stotz2005,Bertrand2016,Takada2019,Delsing2019,Jadot2021,Ito2021,Wang2023}. Recently, research has been conducted towards the realization of electron flying qubits as quantum information processing devices \cite{FQubit2022}.

In electron-quantum optics, single electrons are controlled using basic tools such as a single-electron source and a single-electron detector. These tools have been realized in several different systems \cite{Feve2007,Dubois2013,Jullien2014,Bisognin2019}, including the SAW system. In the SAW system by combining these tools, single-electron transport over micrometers has been demonstrated \cite{Hermelin2011,McNeil2011,Takada2019}, and coherent transport of electron spin has also been demonstrated \cite{Bertrand2016,Jadot2021}.

A typical setup for a single-electron source using SAWs combines a quantum dot (QD), a transport channel, and an interdigital transducer (IDT). A single electron is prepared in the QDs and transported along the depleted transport channel by the SAWs. By using sufficiently strong SAWs, the transport of electrons becomes highly robust, and an electron is transferred while confined to a specific potential minimum of the SAWs \cite{HE_APL2021}. This allows us to control the transfer timing of an electron and hence synchronized transfer of single electrons from multiple single-electron sources is possible. Recently, an electron-collision experiment using two synchronized single-electron sources in the SAW system has been demonstrated \cite{Wang2023}. Another important development is a technique to generate a single-cycle SAW pulse using a chirped-interdigital transducer (a chirp-IDT) \cite{JW_PRX2022}. Employing the SAW pulse for single-electron transfer allows us to synchronize the timing of electron transport from multiple single-electron sources without \SI{}{ps}-triggering each QD \cite{Takada2019}. This is advantageous for scaling up the system since we do not need to implement an RF line for each QD used as a single-electron source. On the other hand, in these previous studies, the single electron ejected had to be prepared in the quantum dot first. Thus, each single-electron source required a QD and a complex fast voltage sequence. This indicates that there is still room for improvement in the scalability of the system. Furthermore, the time-consuming electron preparation process is a speed limiting factor for the entire electron flying qubit system that utilizes single electron charges in flight.

For metrology applications, regular SAWs are used with a depleted quantum rail consisting of two metal gates to realize a quantized current source \cite{Shilton1996,Talyanskii1997,Cunningham1999,Cunningham2000,Ebbecke2002,Kataoka2006}. Electric potential modulations accompanied by the SAWs pick up an integer number of electrons directly from the Fermi sea and transfer them over the quantum rail. As a result, the current, $I = nef$, where $n$ is an integer, $e$ is the elementary charge, and $f$ is the frequency of the SAWs. Here when a single-cycle SAW pulse rather than regular SAWs is used with a depleted quantum rail, an on-demand single-electron source rather than a continuous quantized current source can be realized. In contrast to the single-electron source with a QD, the electron capturing process before applying SAWs is not required. Hence, its operation could be faster. Furthermore, a chirp IDT can generate multiple SAW pulses with arbitrary delays, making it possible to generate single electrons with flexible arbitrary delays. So far, for metrology high accuracy electron pumping (an error rate $\leq 10^{-6}$) is reported in other methods \cite{Keller1996, Stein2015, Yamahata2016, Blumenthal2007}. SAW pumps have not been actively studied for standard application due to its limited accuracy (an error rate $\geq 10^{-4}$), nevertheless their application to quantum information processing may not require the same level of high accuracy as the current standard, and as mentioned above, active research is underway.

Based on this idea, in this paper, we actualize an on-demand single-electron source consisting of a chirp IDT and a quantum rail for enabling the implementation of electron flying qubits with beam-splitters \cite{Ito2021,Wang2023}. We evaluate the performance of this single-electron source by repeatedly operating the single-electron source and by measuring the accuracy of the generated quantized current. In addition, we demonstrate that the delay between the successively transferred single electrons can be arbitrarily adjusted. Finally, we investigate the effect of electromagnetic crosstalk generated by driving the IDT on the accuracy of the single-electron source.

\section{EXPERIMENTAL SETUP}
\begin{figure}
\centering
\includegraphics[width=0.5\textwidth]{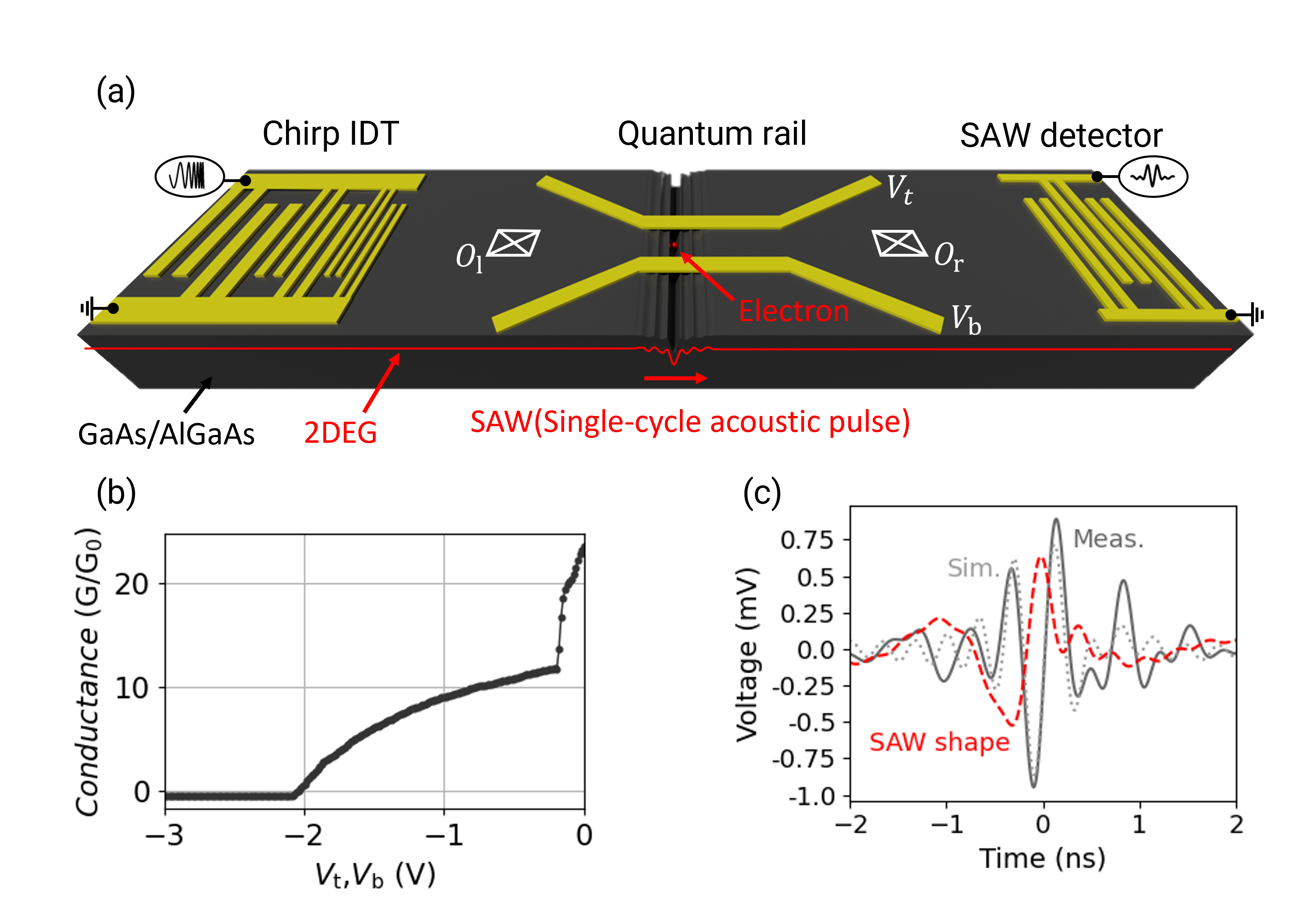}
\caption{\label{fig:Figure1}(a) Experimental setup. Schematic of a chirp IDT emitting a compressed SAW towards a quantum rail and a broadband SAW detector, showing a perspective view of the sample that is realized via a metal surface gate in a GaAs/AlGaAs heterostructure. (b) Conductance across the quantum rail as a function of the voltages $V_{\rm t}$ and $V_{\rm b}$. The current is measured from the ohmic contact $O_{\rm l}$ while applying a DC bias voltage (\SI{336}{\micro V}) to the ohmic contact $O_{\rm r}$. $G_0 = 2e^{2}/h$, where $e$ is the electron charge and $h$ is the Planck constant. (c) Trace of the broadband detector's response to the compressed SAW generated by the chirp IDT (gray solid line) with impulse-response simulation (light gray dotted line) and the corresponding SAW shape (red dashed line) which is derived by deconvolving the detector response to remove the contribution of the detector IDT in the simulation. The measurement is performed at \SI{4}{K}.}
\end{figure}

The experiment is performed within a \SI{4}{K} pulse tube refrigerator. A Si-modulation-doped GaAs/AlGaAs heterostructure is used to fabricate the sample. The two-dimensional electron gas (2DEG) that is located at \SI{100}{nm} below the surface has an electron density of  $n\approx$ \SI{2.8e11}{cm^{-2}} and a mobility of $\mu\approx$ \SI{9e5}{cm^{2}V^{-1}s^{-1}}. Fig.\,\ref{fig:Figure1}a shows a schematic diagram of the device and the experimental setup. The device contains a quantum rail with a lithographic width of \SI{0.8}{\micro m} and a length of \SI{2}{\micro m}, defined by surface Schottky gates. The gates are made of a thin metal film consisting of \SI{3}{nm} titanium and \SI{14}{nm} gold. During the cooldown of the device, a voltage of \SI{0.35}{V} is applied to all the gates. The 2DEG around the gates is depleted by applying negative voltages. Fig.\,\ref{fig:Figure1}b shows a conductance across the quantum rail as a function of the voltage $V_{\rm t}$ and $V_{\rm b}$. For this measurement, we inject a current by applying a DC bias voltage (\SI{336}{\micro V}) to the contact $O_{\rm r}$ and measure the current recovered from the contact $O_{\rm l}$. In the following measurements, we set the voltages $V_{\rm t}$, $V_{\rm b}$ to be more negative than \SI{-2.1}{V}, where the current driven by the bias voltage does not flow. A chirp-IDT is placed on the sample surface \SI{1.4}{mm} to the left of the quantum rail. The surface electrodes of the IDTs are made of a thin metal film consisting of \SI{3}{nm} titanium and \SI{27}{nm} aluminum. To reduce internal reflections at resonance, we employ a double-electrode pattern for the IDT. The IDT aperture is \SI{30}{\micro m}, and the SAW propagation direction is along [110]. The IDTs are designed and simulated with the homemade open-source Python library “idtpy” \cite{idtpy}. A signal for generating SAW is produced by an arbitrary waveform generator (AWG, Keysight M8190A). This signal subsequently passes through two high-frequency amplifiers (SHF S126A and ZHL-4W-422+) at room temperature before being inputted into the chirp IDT. The generated SAW can be observed by a high-speed sampling oscilloscope (Keysight N1094B DCA-M) via the broadband detector IDT after being amplified by a broadband amplifier (SHF S126A) at room temperature.

\section{SAW PULSE GENERATION}
A chirp IDT \cite{JW_PRX2022} has a cell periodicity that changes gradually as shown in Fig.\,\ref{fig:Figure1}a.
Here it is designed to generate SAWs with frequencies ranging from \SI{0.5}{GHz} to \SI{3.0}{GHz}. 
By applying an appropriately time-dependent high-frequency voltage to this IDT, it is possible to generate strongly compressed SAW pulses. 
The gray solid line in Fig.\,\ref{fig:Figure1}c is a strongly compressed SAW pulse observed by the detector IDT. 
This waveform is distorted from the actual shape of the SAW that passes through the quantum rail, due to the frequency bandwidth of the detector. 
To find the waveform of the SAW transporting electrons, we performed a simulation using the impulse-response model. 
First, we simulated the waveform of the SAW including the detector-IDT component (light gray dotted line in Fig.\,\ref{fig:Figure1}c). 
Then, by subtracting the detector-IDT component, the waveform of the SAW in the device was simulated (red dashed line in Fig.\,\ref{fig:Figure1}c). The result indicates that a SAW pulse with one dominant minimum can be generated. 
What we would like to focus on here is the shape of the SAW pulse. This shape is optimized to smoothly pick up single electrons from the Fermi sea and transport them across the depleted quantum rail, which differs from the shape used for the purpose of single electron transport using the QD. As a result of optimization, an asymmetric SAW pulse with a sharp edge after the minimum value was obtained.

\section{ON-DEMAND SINGLE ELECTRON SOURCE WITH SAW PULSES}
It has been demonstrated that acousto-electric currents can be generated by applying SAWs to a depleted quantum rail \cite{Shilton1996}.
Here the superposition of the longitudinal dynamic potential of the SAW and the transversal confinement potential of the quantum rail forms a train of moving QDs that picks up electrons from the Fermi sea and carries them across.
The average number of electrons carried in each dynamic QD is determined at the entrance of the quantum rail by the balance between the potential gradient
towards the entrance of the quantum rail and the confinement potential of the moving potential minima.
When the potential gradient of the dynamic QD becomes steeper due to changes in the voltage of the quantum rail or the waveform profile of the SAW, the spacing between the energy levels of the electrons in the QD widens and the charging energy of the QD increases.
For a sufficiently large charging energy, the number of electrons within each dynamic QD is stably quantized. 
When each dynamic QD contains $n$ electrons, where $n$ is an integer number, the device works as a continuous quantized-electron source and generates a quantized current, $I=nef$, where $f$ is the frequency of the sinusoidal SAW.
Here we investigate such a quantized current source with the single-cycle acoustic pulse shown above. 
By repeatedly sending the SAW pulse to the depleted quantum rail we generate an observable quantized current and evaluate the accuracy of the single-electron transport by each SAW pulse from the stability of the current quantization. In this experiment, we set the repetition period of the SAW pulse to $T_{\rm cycle} = \SI{1280}{ns}$. When the number of electrons transported by each SAW pulse is quantized to an integer number $n$, the quantized current, $I=ne/T_{\rm cycle}$, is expected to be observed. Since we can arbitrarily control the timing of the SAW-pulse generation with a chirp IDT, this electron source can be considered as an on-demand quantized electron source.

\begin{figure}
\centering
\includegraphics[width=0.5\textwidth]{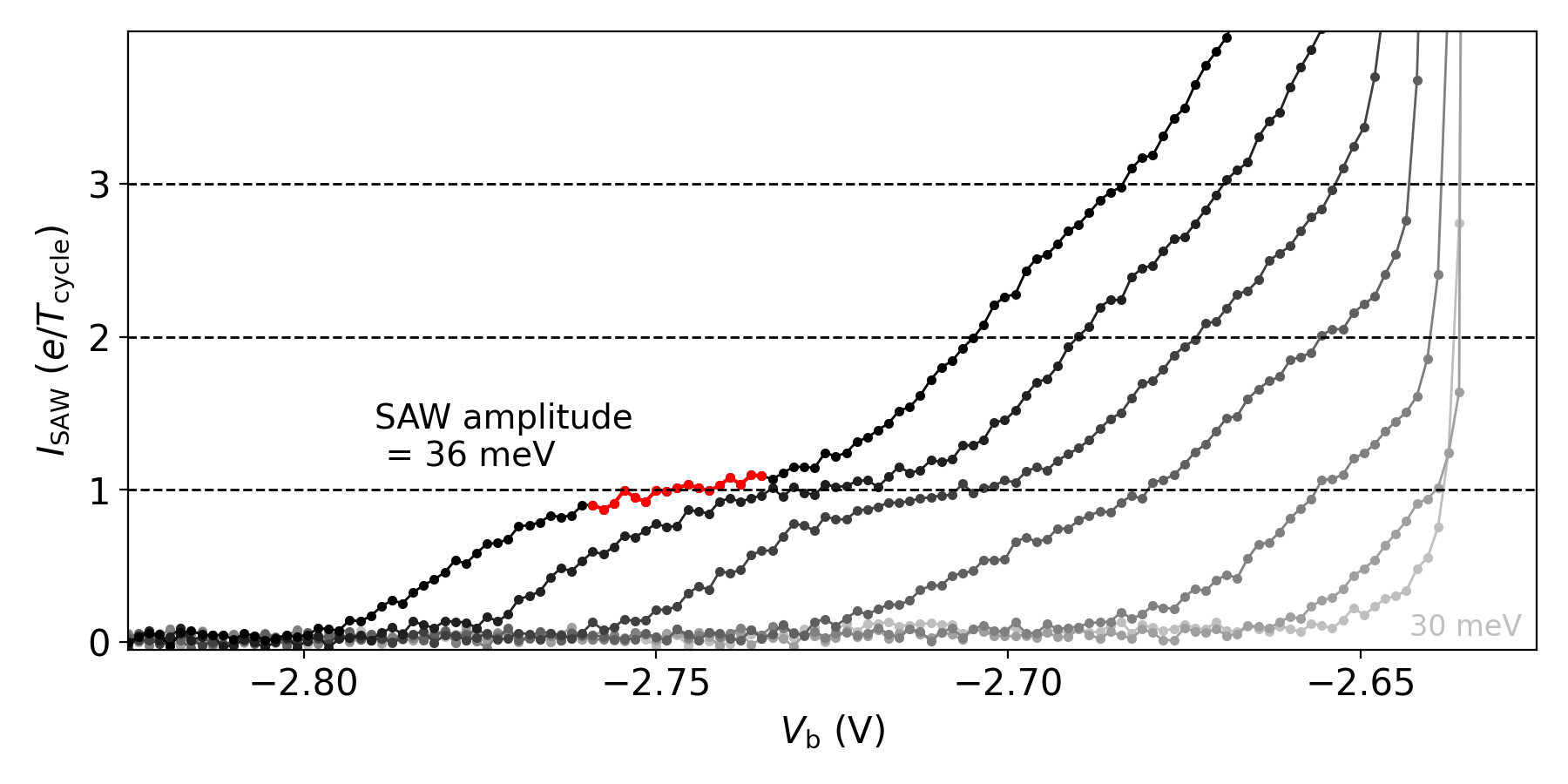}
\caption{\label{fig:Figure2}Acousto-electric current, $I_{\rm SAW}$, induced by the compressed SAW pulse as a function of the voltage $V_{\rm b}$ with $V_{\rm u}=\SI{-2.2}{V}$. The SAW amplitude varies from \SI{30}{meV} to \SI{36}{meV} (from right to left) (refer to Appendix B). The range indicated in red is a flat region where the gradient is less than a certain value of the most left curve (refer to Appendix A).}
\end{figure}

Fig.\,\ref{fig:Figure2} displays the acousto-electric current as a function of the gate voltage of the quantum rail for different SAW amplitudes. 
When the gate voltage is swept to a more negative value, the potential gradient at the entrance of the quantum rail increases. 
As a result, a smaller number of electrons is transported across the depleted quantum rail by the SAW-dynamical potential.
For the smaller SAW amplitude, the potential gradient of the SAW-dynamical potential is smaller and the charging energy of the dynamic QD at the entrance of the quantum rail is not large enough to have a stable number of electrons in each dynamic QD.
This results in the acousto-electric current smoothly decreasing as a function of the gate voltage.
In contrast, for the larger SAW amplitude, the charging energy increases, and a kink is developed at $e/T_{\rm cycle}$.
To evaluate the quantization of the acoustic-electric current, we focus on the region where the gradient of the kink is the flattest as indicated in red dots in Fig.\,\ref{fig:Figure2}. We calculated the normalized difference, $$I_{\rm N}=|(I_{\rm ave}-e/T_{\rm cycle})/(e/T_{\rm cycle})|,$$ between the average of measured acousto-electric current $I_{\rm ave}$ and ideal estimated value $e/T_{\rm cycle}$ with a combined standard uncertainty. The acoustic-electric current took into account the current without SAW as offset and the gain of the current amplifier calibrated by a standard resistance. In the flattest region (the gate voltage width of \SI{24}{mV}), the normalized difference reaches below ($0.037 \pm 0.013$) (refer to Appendix A). This indicates that the kink caused by the SAW admirably matched the ideal quantized current, suggesting that in the kink, the number of electrons in each SAW pulse is mostly quantized to one. Hence, the realization of an on-demand single electron source has been achieved.

In this method, single electrons can be generated without preparing electrons, offering faster operation than conventional on-demand single-electron sources using SAWs. Additionally, the interval of SAW pulses can be flexibly controlled, making it possible to generate single electrons with any desired delay as discussed in the following section.

\section{SINGLE-ELECTRON SOURCE WITH AN ARBITRARY DELAY}
\begin{figure}
\centering
\includegraphics[width=0.5\textwidth]{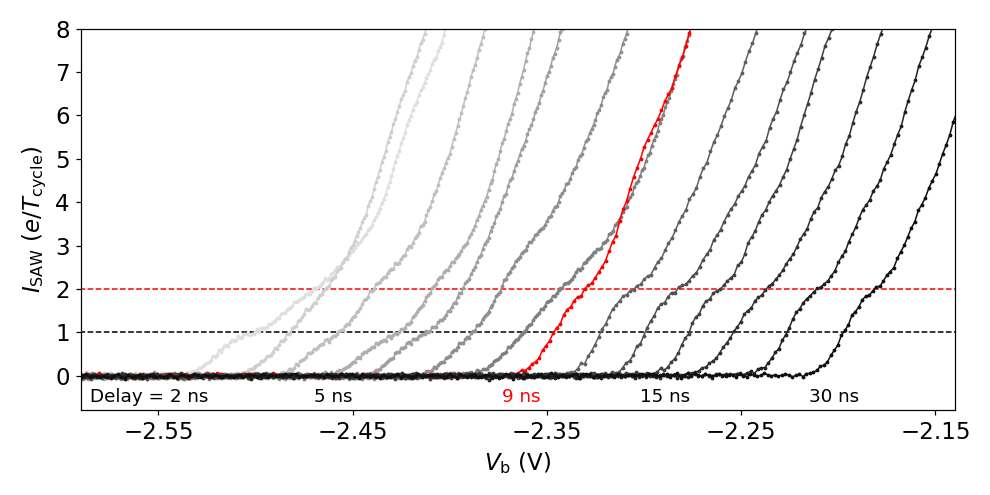}
\caption{\label{fig:Figure3}
 Acousto-electric current, $I_{\rm SAW}$, as a function of the voltage $V_{\rm b}$ with offset for clearly \cite{OffsetFig3}. Two SAW pulses within $T_{\rm cycle}$ with changing delay between the pulses from \SI{2}{ns} to \SI{30}{ns} (from left to right). }
\end{figure}

In the previous section, the interval between successive SAW pulses was set to \SI{1280}{ns}, which is longer than the length of the SAW generation signal of \SI{130}{ns}. By setting the interval longer, the SAW generation signals did not overlap each other. However, it is also possible to generate SAW pulses with shorter intervals by purposely overlapping the SAW generation signals. In this section, we use this technique to verify the operation of a single electron pump with arbitrarily controlled delay and explore its potential for high-speed operation.
One limitation to keep in mind is the maximum output power of the high-frequency amplifier used: when two SAW generation signals overlap, the maximum amplitude of the individual SAW pulses only reaches half the normal amplitude. This limitation reduces the flatness of the quantized current kink. Nevertheless, despite this reduced flatness, our data still exhibit distinguishable characteristics of a single-electron pump as the kinks seen in Fig.\,\ref{fig:Figure3}. Fig.\,\ref{fig:Figure3} presents the acousto-electric current as a function of the gate voltages when we control the delay between two successive SAW pulses within $T_{\rm cycle}$ between \SI{2}{ns} and \SI{30}{ns}. The kinks will appear at $2e/T_{\rm cycle}$ since we send two SAW pulses within $T_{\rm cycle}$.
For the delays shorter than \SI{9}{ns} the kinks appear at unstable positions.
Whereas, for the delays larger than \SI{9}{ns} the $2e/T_{\rm cycle}$ quantization current is stably observed.
We attribute this result to the presence of small acoustic fluctuation before and after the main acoustic minimum. These extra fluctuations of the SAW pulse overlap with the main minimum of the other SAW pulse, preventing stable electron generation. When a chirp IDT with a wider frequency bandwidth is developed, such extra fluctuation can be suppressed, and a shorter delay time than \SI{9}{ns} would be possible. In principle, the delay time can be shortened down to the width of the primary minimum, which is approximately \SI{1}{ns} in this study. In our current chirp IDT, the delay can be arbitrarily controlled above \SI{9}{ns}.

\section{EFFECT OF ELECTROMAGNETIC CROSSTALK}
For the stable electron-pump operation, the influence of electromagnetic crosstalk has to be taken into account. In the process of exciting an IDT to generate a SAW, an electromagnetic wave is concurrently emitted from the IDT. This emanation disrupts the surrounding potential of the nanostructures such as the quantum rail, thereby impeding the stability of electron-pump operations. In previous single-electron transfer experiments with SAWs, the difference in velocity between SAWs ($v_\mathrm{SAW} = \SI{2.81}{\micro m/ns}$ \cite{HE_APL2021}) and electromagnetic waves was used to avoid the simultaneous arrival of SAWs and electromagnetic waves to the nanostructures and to suppress the influence of such crosstalk.
The influence of electromagnetic crosstalk on a single-electron pump using a standard IDT has been discussed and pointed out as an important problem previously \cite{Kataoka2006}. There, the crosstalk was suppressed by the pulsed operation of the IDT and by avoiding the simultaneous arrival of SAWs and electromagnetic waves. When a standard IDT with a single resonant frequency is excited by pulsed sinusoidal waves, a narrow frequency bandwidth of the standard IDT results in a finite rise (fall) time of SAWs, where a gradual change of the SAW amplitude makes single-electron pump operation unstable. As a result, it is not possible to avoid the influence of electromagnetic crosstalk with a standard IDT while keeping a highly accurate single-electron pump operation.

\begin{figure}
\centering
\includegraphics[width=0.5\textwidth]{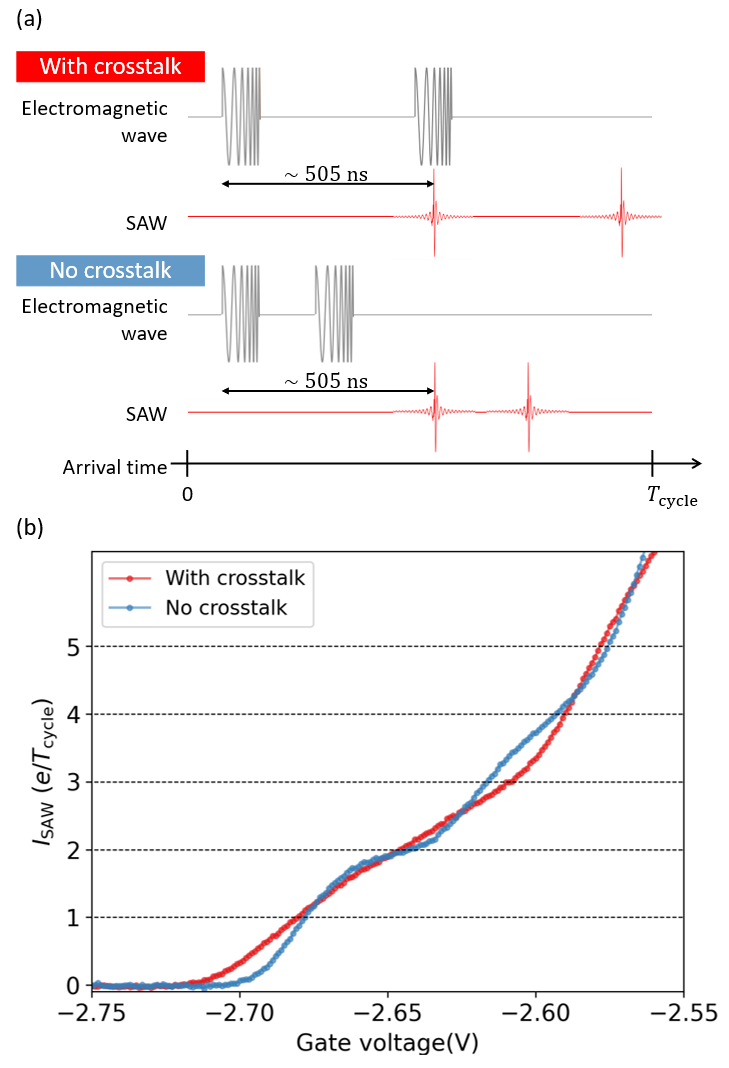}
\caption{\label{fig:Figure4}(a)Schematic of the arrival time of electromagnetic waves and SAW pulses at the quantum rail. The SAW arrives at the quantum rail approximately \SI{505}{ns} after generation at the IDT. (b)Acousto-electric current, $I_{\rm SAW}$, as a function of the voltage $V_{\rm b}$ with $V_{\rm u}=\SI{-2.06}{V}$ with and without crosstalk.}
\end{figure}

On the other hand, the SAW pulses generated by our chirp IDT do not have a rise (fall) time but only a single-potential minimum transporting electrons.
Therefore, we can arbitrarily switch on and off the driving of the chirp IDT without degrading the stability of electron-pump operations.
In the present device, from the distance between the IDT and the quantum rail, the SAW reaches the quantum rail approximately \SI{505}{ns} after its generation at the IDT.
The electromagnetic crosstalk propagates with a velocity of light and reaches the quantum rail immediately after its generation at the IDT.
The influence of the crosstalk can be suppressed by shifting the timing of the SAW pulse arrival at the quantum rail and the timing of the SAW generation signal input to the IDT as shown in Fig.\,\ref{fig:Figure4}a. Fig.\,\ref{fig:Figure4}b shows the influence of the electromagnetic crosstalk on acousto-electric transport.
Here, the number of SAW pulses within one cycle $T_{\rm cycle}$ is fixed at 2, changing only the timing of the SAW pulses. As a result, for one condition (red curve in Fig.\,\ref{fig:Figure4}b) the SAW pulses and electromagnetic waves reach the quantum rail at the same time and hence the crosstalk effect exists.
For the other case (blue curve in Fig.\,\ref{fig:Figure4}b), by avoiding the simultaneous arrival the crosstalk effect is suppressed. When the crosstalk effect exists, stable electron pumping is disturbed and acousto-electric current smoothly changes as a function of the gate voltages.
On the other hand, when the crosstalk effect is properly suppressed, a kink appears at the expected value $2e/T_{\rm cycle}$.

\section{SUMMARY AND OUTLOOK}
In essence, we have demonstrated a simple on-demand single-electron source amalgamating chirp SAW pulses with a quantum rail.
We evaluated its performance from the observation of the quantized acousto-electric current generated by repeatedly operating the source. Under the optimal operation condition, the mismatch of $3.7 \%$ or less compared to the ideal operation.
This single-electron source negates the need for dynamic gate operations to prepare a single electron into the QD, which stands in stark contrast to the formerly demonstrated single-electron source with chirp SAW pulses \cite{JW_PRX2022}. 
Furthermore, we demonstrated the flexible control of a delay between successive single-electron transfers. Meanwhile, the shortest delay time is limited to \SI{9}{ns} and the operation accuracy is limited by the maximum available SAW amplitude.
The former limitation could potentially be overcome by expanding the bandwidth of a chirp IDT. The latter can be ameliorated by enhancing the conversion efficiency between IDT input signals and SAWs, achievable through the utilization of a thin film of stronger piezoelectric materials than GaAs such as ZnO or AlN, or by fine-tuning the impedance mismatch in the IDT. With the improvements the accuracy of the single-electron pump operation will also be enhanced. As the width of the quantum rail (\SI{0.8}{\micro m} in this work) is much narrower than the wavefront of the SAW ($\sim$ \SI{30}{\micro m} in this work and can be even wider), synchronized operations of multiple single-electron sources can be implemented by simply putting multiple quantum rails within the wavefront of a SAW. Notably, each quantum rail can be implemented with at least two or fewer static voltage inputs and does not require complex voltage manipulation. These characteristics facilitate the integration of many parallel single-electron sources and encourage scale-up of electron flying qubit architectures.

Another important insight gained from this work is the impact of the SAW potential shape on electron transport. A SAW pulse, generated by a chirp IDT, is a superposition of broadband SAWs, thus permitting the deformation of the SAW pulse shape through appropriate adjustment of the input signal to the IDT. This study required significant deformation of the SAW pulse into an optimized asymmetric shape, in order to directly extract an electron from the Fermi sea instead of the QD. It indicates that the previously used symmetric SAW waveform was not the most suitable for electron transport and suggests the direction for further optimization. This insight is not only beneficial for research using SAW but also has important implications for studies on the electron transport process in nanostructures \cite{Langrock2023,Seidler2022}.

Additionally, we explored the effect of the electromagnetic field emitted directly from the IDT on the accuracy of the single-electron source. It has been presented as a factor to degrade the accuracy of electron transfer and pulse modulation of SAWs has been proposed as a solution for that \cite{Kataoka2006}. However, the narrow bandwidth of the IDT prevented the fast enough pulse modulation to eliminate the influence of the electromagnetic crosstalk. In contrast, the large bandwidth of our chirp IDT and a single-cycle SAW pulse originating from it allow us to completely separate the timing of single-electron transfer across the quantum rail and the arrival of the electromagnetic crosstalk. We clearly demonstrate that the elimination of the crosstalk indeed improves the accuracy of electron transport.

The results obtained in this study gives important insights into the single-electron transport with moving electric potentials and contribute to the field of single-electron quantum optics using SAWs, such as building up a flying qubit system or quantum communication with single electron (or hole) to single photon conversion \cite{Hsiao2020}. This study represents steady progress towards the realization of quantum systems using single electrons, providing new techniques and insights that enrich the fundamental understanding of the field.

\section*{ACKNOWLEDGMENTS}
S.O. acknowledge financial support from JST SPRING, Grant Number JPMJSP2106.\\
J.W. acknowledges the European Union's Horizon 2020 research and innovation program under the Marie Skłodowska-Curie grant agreement No 754303. \\
T.K. and S.T. acknowledge financial support from JSPS KAKENHI Grant Number 20H02559. \\
N.-H.K. acknowledges financial support from JSPS KAKENHI Grant Number JP18H05258. \\
C.B. acknowledges financial support from
the French Agence Nationale de la Recherche (ANR), project QUABS ANR-21-CE47-0013-01 as well as funding from the European Union’s H2020 research and innovation program under grant agreement No 862683 ”UltraFastNano”.\\

\section*{Appendix A: UNCERTAINTY OF THE ACOUSTO-ELECTRIC CURRENT}
\begin{figure}
\centering
\includegraphics[width=0.5\textwidth]{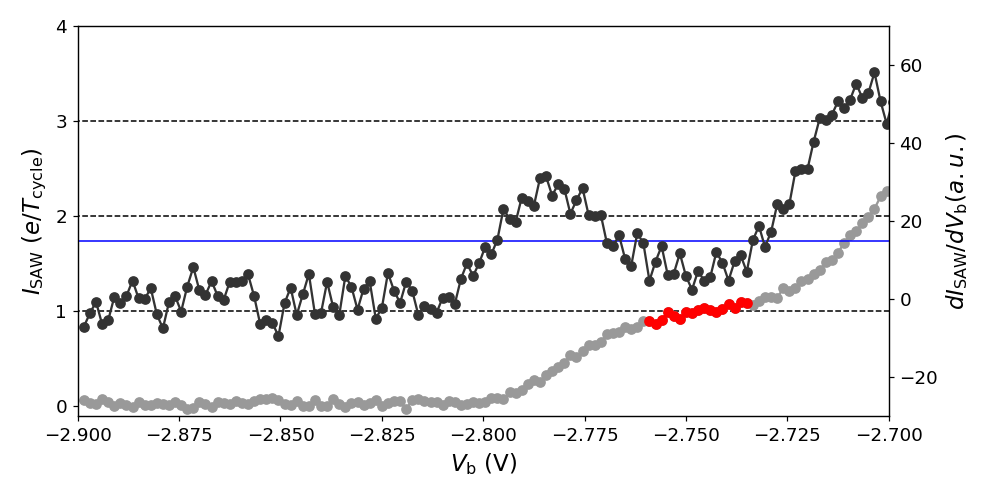}
\caption{\label{fig:Figure5} Acousto-electric current, $I_{\rm SAW}$, of maximum SAW amplitudes in Fig.\,\ref{fig:Figure2} and its gradient. The left y-axis represents $I_{\rm SAW}$ (gray and red), and the right y-axis represents the gradient of $I_{\rm SAW}$ (black). The gradient is smoothed by averaging the values of the five neighboring points.}
\end{figure}

The normalized difference,
$$
I_{\rm N}=|(I_{\rm ave}-e/T_{\rm cycle})/(e/T_{\rm cycle})|,
$$
was calculated from the average of measured acousto-electric current $I_{\rm ave}$ and ideal estimated value $e/T_{\rm cycle}$. The average of current value $I_{\rm ave}$ and the combined standard uncertainty were derived as follows: $I_{\rm ave}$ is the average of measured values within the range in which the gradient becomes small and maintains a certain level of constancy. The range of gradient smaller than -15 (blue solid line in Fig.\,\ref{fig:Figure5}) was selected for our analysis as indicated in the red region in Fig.\,\ref{fig:Figure5}. As an offset, the average of measured values where the zero current flow is subtracted, and the uncertainty of the trans-impedance amplifier (DDPCA-300) is also factored in. The gain of the trans-impedance amplifier was calibrated using high-ohm standard resistors calibrated by national resistance standards. We utilized a combined standard uncertainty that included the gain uncertainty of the trans-impedance amplifier and the standard uncertainty derived from the current measurements.

\section*{Appendix B: SAW AMPLITUDE ESTIMATION}
The amplitude of the SAW when using chirp pulses has been estimated in a previous report \cite{JW_PRX2022} by comparison with the SAW stemming from the normal IDT. This previous report and the experiment in this paper were performed with the same setup and the same substrate. From the power-to-energy conversions that have been obtained, the amplitude of the compressed SAW pulse generated by the input power to the chirp IDT of \SI{28.6}{dBm} (\SI{30.1}{dBm}) sent from room temperature to the cryogenic setup is estimated to be \SI{30}{meV} (\SI{36}{meV}) in data of Fig.\, \ref{fig:Figure2}. We mainly used SAWs with an amplitude of \SI{30}{meV} in other measurements. In Fig.\,\ref{fig:Figure3} each SAW pulse is half the amplitude as a result of overlapping input waveforms, so \SI{15}{meV}. Because the condition of the quantum rail changes due to the recooling of the sample, the gate voltage condition at which the kink appears is different in the results of each figure, even though the SAW amplitude is the same for each measurement.

\bibliographystyle{unsrtnat}
\bibliography{template}  %%% Uncomment this line and comment out the ``thebibliography'' section below to use the external .bib file (using bibtex) .

%%% Uncomment this section and comment out the \bibliography{references} line above to use inline references.
% \begin{thebibliography}{1}

% 	\bibitem{kour2014real}
% 	George Kour and Raid Saabne.
% 	\newblock Real-time segmentation of on-line handwritten arabic script.
% 	\newblock In {\em Frontiers in Handwriting Recognition (ICFHR), 2014 14th
% 			International Conference on}, pages 417--422. IEEE, 2014.

% 	\bibitem{kour2014fast}
% 	George Kour and Raid Saabne.
% 	\newblock Fast classification of handwritten on-line arabic characters.
% 	\newblock In {\em Soft Computing and Pattern Recognition (SoCPaR), 2014 6th
% 			International Conference of}, pages 312--318. IEEE, 2014.

% 	\bibitem{hadash2018estimate}
% 	Guy Hadash, Einat Kermany, Boaz Carmeli, Ofer Lavi, George Kour, and Alon
% 	Jacovi.
% 	\newblock Estimate and replace: A novel approach to integrating deep neural
% 	networks with existing applications.
% 	\newblock {\em arXiv preprint arXiv:1804.09028}, 2018.

% \end{thebibliography}

\end{document}